# Detect Adverse Drug Reactions for Drug Aspirin

Yihui liu and Uwe Aickelin

*Abstract*—**Adverse drug reaction (ADR) is widely concerned for public health issue. In this study we propose an original approach to detect the ADRs using feature matrix and feature selection. The experiments are carried out on the drug Aspirin. Major side effects for the drug are detected and better performance is achieved compared to other computerized methods. The detected ADRs are based on the computerized method, further investigation is needed.**

## I. INTRODUCTION

ADVERSE drug reaction (ADR) is widely concerned for public health issue. ADRs are one of most common causes to withdraw some drugs from market [1]. Now two major methods for detecting ADRs are spontaneous reporting system (SRS) [2], [3], and prescription event monitoring (PEM)[4],[5]. The World Health Organization (WHO) defines a signal in pharmacovigilance as "any reported information on a possible causal relationship between an adverse event and a drug, the relationship being unknown or incompletely documented previously" [6]. For spontaneous reporting system, many machine learning methods are used to detect ADRs, such as Bayesian confidence propagation neural network (BCPNN) [7], decision support method [8], genetic algorithm [9], knowledge based approach [10], etc. One limitation is the reporting mechanism to submit ADR reports, which has serious underreporting and is not able to accurately quantify the corresponding risk. Another limitation is hard to detect ADRs with small number of occurrences of each drug-event association in the database.

In this paper we propose feature selection approach to detect ADRs from The Health Improvement Network (THIN) database. First feature matrix, which represents the medical events for the patients before and after taking drugs, is created by linking patients' prescriptions and corresponding medical events together. Then significant features are selected based on feature selection methods, comparing the feature matrix before patients take drugs with one after patients take drugs. Finally the significant ADRs can be detected from thousands of medical events based on corresponding features. Experiments are carried out on the drug Aspirin. Good performance is achieved.

## II. FEATURE MATRIX AND FEATURE SELECTION

### A. The Extraction Of Feature Matrix

To detect the ADRs of drugs, first feature matrix is extracted from THIN database, which describes the medical events that patients occur before or after taking drugs. Then feature selection method of Student's t-test is performed to select the significant features from feature matrix containing thousands of medical events. Figure 1 shows the process to detect the ADRs using feature matrix. Feature matrix $A$ describes the medical events for each patient during 60 days before they take drugs. Feature matrix $B$ reflects the medical events during 60 days after patients take drugs. In order to reduce the effect of the small events, and save the computation time and space, we set 100 patients as a group. Matrix $X$ and $Y$ are feature matrix after patients are divided into groups.

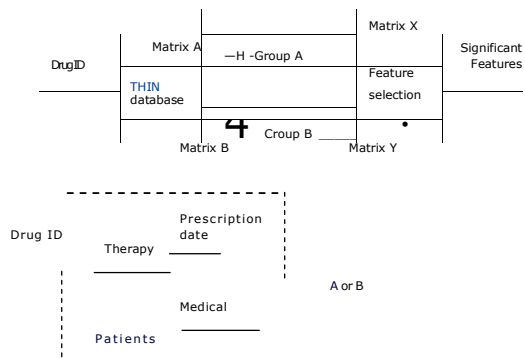

Fig. 1. The process to detect ADRs. Matrix $A$ and $B$ are feature matrix before patients take drugs or after patients take drugs. The time period of observation is set to 60 days. Matrix $X$ and $Y$ are feature matrix after patients are divided into groups. We set 100 patients as one group.

### B. Medical Events And Readcodes

Medical events or symptoms are represented by medical codes or Readcodes. There are 103387 types of medical events in "Readcodes" database. The Read Codes used in general practice (GP), were invented and developed by Dr James Read in 1982. The NHS (National Health Service) has expanded the codes to cover all areas of clinical practice. The code is hierarchical from left to right or from level 1 to level 5. It means that it gives more detailed information from level 1 to level 5. Table 1 shows the medical symptoms based on Readcodes at level 3 and at level 5. 'Other soft tissue disorders' is general term using Readcodes at level 3. 'Foot pain', 'Heel pain', etc., give more details using Readcodes at level 5.

Y Liu is with Institute of Intelligent Information Processing, Shandong Polytechnic University, China, 250353, and was with School of Computer Science, University of Nottingham, UK. (corresponding author to provide phone: 0086-531-89631256; fax: 0086-531-89631256; e-mail: yihui_liu2005@yahoo.co.uk).

U Aickelin is with School of Computer Science, University of Nottingham, UK. (e-mail: Uwe.Aickelin@nottingham.ac.uk).

## C. Feature Selection Based On Student's t-test

Feature extraction and feature selection are widely used in biomedical data processing [11]-[18]. In our research we use feature selection method of Student's t-test [19] to detect the significant ADRs from thousands of medical events. Student's t-test is a kind of statistical hypothesis test based on a normal distribution, and is used to measure the difference between two kinds of samples.

TABLE I.
MEDICAL EVENTS BASED ON READCODES AT LEVEL 3 AND LEVEL 5.

| | Level Readcodes | Medical events |
|---|---|---|
| Muscle pain | Level 3 | N24..00 | Other soft tissue disorders |
| | | N245.16 | Leg pain |
| | | N245111 | Toe pain |
| | Level 5 | N245.13 | Foot pain |
| | | N245700 | Shoulder pain |
| | | N245.15 | Heel pain |

## D. Other Parameters

The variable of ratio $R$, is defined to evaluate significant changes of the medical events, using ratio of the patient number after taking the drug to one before taking the drug. The variable $R_2$ represents the ratio of patient number after taking the drug to the number of whole population having one particular medical symptom.

The ratio variables $R_1$ and $R_2$ are defined as follows:

$$R_1 = \begin{cases} N_A / N_B & \text{if } N_B \neq 0; \\ N_A & \text{if } N_B = 0; \end{cases}$$

$$R_2 = 100 * N_A / N$$

where $N_B$ and $N_A$ represent the numbers of patients before or after they take drugs for having one particular medical event respectively. The variable $N$ represents the number of whole population who take drugs.

## III. EXPERIMENTS AND RESULTS

Drugs.com provides the side effects of Aspirin in [20], [21], including gastrointestinal, renal, hematologic, hypersensitivity, dermatologic, hepatic, oncologic, metabolic, cardiovascular, nervous system, musculoskeletal, respiratory, endocrine, ocular side effects. Some symptoms are heartburn, nausea, upset stomach, severe allergic reactions (rash; hives; itching; difficulty breathing; tightness in the chest; swelling of the mouth, face, lips, or tongue), black or bloody stools, confusion, diarrhea, dizziness, drowsiness, hearing loss, ringing in the ears, severe or persistent stomach pain, unusual bruising, vomiting, etc.

17125 patients are taking Aspirin (ASPIRIN disp tab 75mg, and 300mg) from 20GPs data in THIN database, removing the patients registered in GP less than 12 months. Based on Readcodes at level 1-5, totally 14546 medical events are obtained before or after patients take the drug. So 17125x14546 feature matrix is obtained. Based on Readcodes at level 1-3, we combine the medical events, which have the same first three codes, into one medical event. Totally 2853 medical events are obtained, and 17125x2853 feature matrix are created. After grouping them, 171x14546 and 171x2853 feature matrix are formed to select the significant features, which reflect the significant change of medical events.

Table 2 shows the top detected results in ascending order of p value of Student's t-test, using Readcodes at level 1-5 and at level 1-3. The detected results are using p value less than 0.05, which represent the significant change after patients take the drug. Table 3 shows the results in descending order of the ratio of the number of patients after taking the drug to one before taking the drug. Table 4 shows potential ADRs related to cancer for Aspirin based on p value of Student's t-test. The detected ADRs are based on the computerized method, further investigation is needed.

## IV. CONCLUSIONS

In this study we propose a novel method to successfully detect the ADRs using feature matrix and feature selection. A feature matrix, which characterizes the medical events before patients take drugs or after patients take drugs, is created from THIN database. The feature selection method of Student's t-test is used to detect the significant features from thousands of medical events. The significant ADRs, which are corresponding to significant features, are detected. Experiments are performed on the drug Aspirin. The detected ADRs are based on the computerized method, further investigation is needed.

REFERENCES

[1] G. Severino and M.D. Zompo, "Adverse drug reactions: role of pharmacogenomics," *Pharmacological Research*, vol. 49, pp. 363-373, 2004.
[2] Y. Qian, X Ye, W. Du, J. Ren, Y. Sun, H. Wang, B. Luo, Q. Gao, M. Wu, and J. He, "A computerized system for signal detection in spontaneous reporting system of Shanghai China," *Pharmacoepidemiology and Drug Safety,* vol. 18, pp. 154-158, 2009.
[3] K.S. Park and 0. Kwon, "The state of adverse event reporting and signal generation of dietary supplements in Korea," *Regulatory Toxicology and Pharmacology, vol.* 57, pp. 74-77, 2010.
[4] R. Kasliwal, L.V. Wilton, V. Cornelius, B. Aurich-Barrera, and S.A.W. Shakir, "Safety profile of Rosuvastatin-results of a prescription-event monitoring study of 11680 patients," *Drug Safety,* vol. 30, pp. 157-170, 2007.
[5] R.D. Mann, K. Kubota, G. Pearce, and L. Wilton, "Salmeterol: a study by prescription-event monitoring in a UK cohort of 15,407 patients," *J Clin Epidemiol,* vol. 49, pp. 247-250, 1996..
[6] R.H. Meyboom, M. Lindquist, A.C. Egberts, and I.R. Edwards, "Signal detection and follow-up in pharmacovigilance," *Drug Saf,* vol. 25, pp. 459-465, 2002.
[7] A. Bate, M. Lindquist, I.R. Edwards, S. Olsson, R. Orre, A. Lansner, and R.M. De Freitas, "A Bayesian neural network method for adverse drug reaction signal generation," *Eur J Clin Pharmacol,* vol. 54, pp. 315-321, 1998.
[8] M. Hauben and A. Bate, "Decision support methods for the detection of adverse events in post-marketing data," *Drug Discovery Today,* vol. 14, pp. 343-357, 2009.
[9] Y. Koh, C.W. Yap, and S.C. Li, "A quantitative approach of using genetic algorithm in designing a probability scoring system of an adverse drug reaction assessment system," International Journal of Medical Informatics, vol. 77, pp. 421-430, 2008.
[10] C. Henegar, C. Bousquet, A.L. Lillo-Le, P. Degoulet, and M.C. Jaulent, "A knowledge based approach for automated signal generation in pharmacovigilance," *Stud Health Technol Inform,* vol. 107, pp. 626-630, 2004.
[11] Y. Liu, "Feature extraction and dimensionality reduction for mass spectrometry data," *Computers in Biology and Medicine*, vol. 39, pp. 818-823, 2009.

[12] Y. Liu, "Detect key genes information in classification of microarray data," *EURASIP Journal on Advances in Signal Processing,* 2008. Available: doi:10.1155/2008/612397.
[13] Y. Liu and L. Bai, "Find significant gene information based on changing points of microarray data," *IEEE Transactions on Biomedical Engineering,* vol. 56, pp. 1108-1116, 2009.
[14] Y. Liu, M. Muftah, T. Das, L. Bai, K. Robson, and D. Auer, "Classification of MR Tumor Images Based on Gabor Wavelet Analysis," *Journal of Medical and Biological Engineering,* vol. 32, pp. 22-28, 2012.
[15] Y. Liu, "Dimensionality reduction and main component extraction of mass spectrometry cancer data," *Knowledge-Based Systems,* vol. 26, pp. 207-215, 2012.
[16] Y. Liu, "Wavelet feature extraction for high-dimensional microarray data," *Neurocomputing,* vol. 72, pp. 985-990, 2009.
[17] Y. Liu, W. Aubrey, K. Martin, A. Sparkes, C. Lu, and R. King, "The analysis of yeast cell morphology features in exponential and stationary Phase," *Journal of Biological Systems,* vol. 19, pp. 561-575, 2011.
[18] Y. Liu, "Prominent feature selection of microarray data," *Progress in Natural Science,* vol. 19, pp. 1365-1371, 2009
[19] E. Kreyszig, Introductory Mathematical Statistics, John Wiley, New York, 1970.
[20] http://en.wikipedia.org/wiki/Aspirin.
[21] http://www. drugs. com/sfx/Aspirin-side-effects.html

TABLE II THE TOP POTENTIAL ADRS FOR ASPIRIN BASED ON PVALUE OF STUDENT'S T-TEST.

| | Rank | Readcodes | Medical events | NB | NA | R1 | R2 |
|---|---|---|---|---|---|---|---|
| | 1 | 1Z12.00 | Chronic kidney disease stage 3 | 177 | 1169 | 6.60 | 6.83 |
| | 2 | M03z000 | Cellulitis NOS | 178 | 711 | 3.99 | 4.15 |
| | 3 | F4C0.00 | Acute conjunctivitis | 200 | 702 | 3.51 | 4.10 |
| | 4 | H06z000 | Chest infection NOS | 339 | 1392 | 4.11 | 8.13 |
| | 5 | N131.00 | Cervicalgia - pain in neck | 206 | 660 | 3.20 | 3.85 |
| Level 1-5 | 6 | K197.00 | Haematuria | 55 | 304 | 5.53 | 1.78 |
| | 7 | N143.00 | Sciatica | 126 | 409 | 3.25 | 2.39 |
| | 8 | K190.00 | Urinary tract infection, site not specified | 298 | 1010 | 3.39 | 5.90 |
| | 9 | F46..00 | Cataract | 115 | 475 | 4.13 | 2.77 |
| | 10 | C34..00 | Gout | 134 | 421 | 3.14 | 2.46 |
| | 11 | 1M10.00 | Knee pain | 225 | 752 | 3.34 | 4.39 |
| | 12 | 1992.00 | Vomiting | 87 | 419 | 4.82 | 2.45 |
| | 13 | H060.00 | Acute bronchitis | 222 | 791 | 3.56 | 4.62 |
| | 14 | N245.17 | Shoulder pain | 212 | 710 | 3.35 | 4.15 |
| | 15 | M0...00 | Skin and subcutaneous tissue infections | 63 | 287 | 4.56 | 1.68 |
| | 16 | D00..00 | Iron deficiency anaemias | 55 | 273 | 4.96 | 1.59 |
| | 17 | F45..00 | Glaucoma | 64 | 236 | 3.69 | 1.38 |
| | 18 | F4D0.00 | Blepharitis | 59 | 249 | 4.22 | 1.45 |
| | 19 | F501.00 | Infective otitis externa | 105 | 357 | 3.40 | 2.08 |
| | 20 | J510.00 | Diverticulosis | 38 | 195 | 5.13 | 1.14 |
| | 1 | 1Z1..00 | Chronic renal impairment | 206 | 1389 | 6.74 | 8.11 |
| | 2 | H06..00 | Acute bronchitis and bronchiolitis | 944 | 2930 | 3.10 | 17.11 |
| Level 1-3 | 3 | 19F..00 | Diarrhoea symptoms | 290 | 1069 | 3.69 | 6.24 |
| | 4 | M03..00 | Other cellulitis and abscess | 267 | 960 | 3.60 | 5.61 |
| | 5 | J57..00 | Other disorders of intestine | 115 | 465 | 4.04 | 2.72 |
| | 6 | 171..00 | Cough | 842 | 2548 | 3.03 | 14.88 |
| | 7 | K19..00 | Other urethral and urinary tract disorders | 453 | 1584 | 3.50 | 9.25 |
| | 8 | N24..00 | Other soft tissue disorders | 1044 | 2863 | 2.74 | 16.72 |
| | 9 | H05..00 | Other acute upper respiratory infections | 331 | 1236 | 3.73 | 7.22 |
| | 10 | M22..00 | Other dermatoses | 248 | 802 | 3.23 | 4.68 |
| | 11 | N21..00 | Peripheral enthesopathies and allied syndromes | 324 | 1016 | 3.14 | 5.93 |
| | 12 | F4C..00 | Disorders of conjunctiva | 291 | 971 | 3.34 | 5.67 |
| | 13 | 183..00 | Oedema | 337 | 1187 | 3.52 | 6.93 |
| | 14 | F46..00 | Cataract | 148 | 638 | 4.31 | 3.73 |
| | 15 | 199..00 | Vomiting | 111 | 517 | 4.66 | 3.02 |
| | 16 | 1D1..00 | C/O: a general symptom | 424 | 1461 | 3.45 | 8.53 |
| | 17 | 1B8..00 | Eye symptoms | 255 | 832 | 3.26 | 4.86 |
| | 18 | D00..00 | Iron deficiency anaemias | 103 | 433 | 4.20 | 2.53 |
| | 19 | N13..00 | Other cervical disorders | 226 | 711 | 3.15 | 4.15 |
| | 20 | 173..00 | Breathlessness | 761 | 1826 | 2.40 | 10.66 |
| | 21 | N09..00 | Other and unspecified joint disorders | 585 | 1675 | 2.86 | 9.78 |
| | 22 | 1B1..00 | General nervous symptoms | 645 | 1675 | 2.60 | 9.78 |
| | 23 | B33..00 | Other malignant neoplasm of skin | 85 | 334 | 3.93 | 1.95 |
| | 24 | 1C1..00 | Hearing symptoms | 177 | 566 | 3.20 | 3.31 |
| | 25 | 1M1..00 | Pain in lower limb | 257 | 826 | 3.21 | 4.82 |
| | 26 | F59..00 | Hearing loss | 115 | 446 | 3.88 | 2.60 |
| | 27 | J10..00 | Diseases of oesophagus | 205 | 653 | 3.19 | 3.81 |
| | 28 | 19C..00 | Constipation | 311 | 1067 | 3.43 | 6.23 |
| | 29 | J51..00 | Diverticula of intestine | 101 | 388 | 3.84 | 2.27 |
| | 30 | 16C..00 | Backache symptom | 410 | 1331 | 3.25 | 7.77 |

TABLE III. THE TOP POTENTIAL ADRS FOR ASPIRIN BASED ON DESCENDING ORDER OF R1 VALUE.

| Rank | Readcodes | Medical events | NB | NA | R1 | R2 |
|---|---|---|---|---|---|---|
| 1 | 1Z1E.00 | Chronic kidney disease stage 3A without proteinuria | 0 | 38 | 38.00 | 0.22 |
| 2 | F4F2.00 | Epiphora | 1 | 30 | 30.00 | 0.18 |
| 3 | F591.00 | Sensorineural hearing loss | 1 | 27 | 27.00 | 0.16 |
| 4 | SK16000 | Other hip injuries | 1 | 22 | 22.00 | 0.13 |
| 5 | 1A45.12 | Haematuria - symptom | 1 | 20 | 20.00 | 0.12 |
| 6 | F4A0.00 | Corneal ulcer | 0 | 19 | 19.00 | 0.11 |
| 7 | 16J5.00 | Facial swelling | 1 | 19 | 19.00 | 0.11 |
| 8 | M12..12 | Contact eczema | 0 | 18 | 18.00 | 0.11 |

|  | | | | | | | |
|---|---|---|---|---|---|---|---|
|  | 9 | 196B.00 | Painful rectal bleeding | 1 | 18 | 18.00 | 0.11 |
|  | 10 | G835.00 | Infected varicose ulcer | 1 | 17 | 17.00 | 0.10 |
| Level 11 1-5 | 11 | 199..00 | Vomiting | 1 | 17 | 17.00 | 0.10 |
|  | 12 | B33..00 | Other malignant neoplasm of skin | 2 | 32 | 16.00 | 0.19 |
|  | 13 | K241400 | Acute epididymitis | 1 | 16 | 16.00 | 0.09 |
|  | 14 | M07z200 | Infection finger | 1 | 16 | 16.00 | 0.09 |
|  | 15 | K08..00 | Impaired renal function disorder | 0 | 16 | 16.00 | 0.09 |
|  | 16 | G311400 | Worsening angina | 1 | 16 | 16.00 | 0.09 |
|  | 17 | 1Z1G.00 | Chronic kidney disease stage 3B without proteinuria | 0 | 16 | 16.00 | 0.09 |
|  | 18 | F4Kz411 | Red eye NOS | 0 | 16 | 16.00 | 0.09 |
|  | 19 | Eu32z00 | [X]Depressive episode, unspecified | 0 | 16 | 16.00 | 0.09 |
|  | 20 | K120.12 | Renal calculus | 1 | 15 | 15.00 | 0.09 |
|  | 1 | J05..00 | Other dental disease/condition of teeth/supporting structure | 1 | 18 | 18.00 | 0.11 |
|  | 2 | E13..00 | Other nonorganic psychoses | 1 | 16 | 16.00 | 0.09 |
|  | 3 | F5...00 | Diseases of the ear and mastoid process | 0 | 16 | 16.00 | 0.09 |
|  | 4 | K13..00 | Other kidney and ureter disorders | 1 | 16 | 16.00 | 0.09 |
|  | 5 | 1P5..00 | Aggressive behaviour | 1 | 14 | 14.00 | 0.08 |
|  | 6 | SA3..00 | Open wound of toe(s) | 1 | 14 | 14.00 | 0.08 |
|  | 7 | S83..00 | Other open wound of head | 2 | 26 | 13.00 | 0.15 |
| Level 8 1-3 | 8 | F4A..00 | Keratitis | 4 | 48 | 12.00 | 0.28 |
|  | 9 | G72..00 | Other aneurysm | 0 | 12 | 12.00 | 0.07 |
|  | 10 | SH9..00 | Bum - unspecified | 0 | 12 | 12.00 | 0.07 |
|  | 11 | S60..00 | Concussion | 1 | 12 | 12.00 | 0.07 |
|  | 12 | K....00 | Genitourinary system diseases | 0 | 12 | 12.00 | 0.07 |
|  | 13 | C2...00 | Nutritional deficiencies | 1 | 12 | 12.00 | 0.07 |
|  | 14 | B....00 | Neoplasms | 0 | 11 | 11.00 | 0.06 |
|  | 15 | 157..00 | H/O: menstrual disorder | 0 | 11 | 11.00 | 0.06 |
|  | 16 | K30..00 | Benign mammary dysplasia | 1 | 11 | 11.00 | 0.06 |
|  | 17 | 1A6..00 | Urethral discharge symptom | 1 | 11 | 11.00 | 0.06 |
|  | 18 | J11..00 | Gastric ulcer - (GU) | 7 | 74 | 10.57 | 0.43 |
|  | 19 | K08..00 | Impaired renal function disorder | 2 | 21 | 10.50 | 0.12 |
|  | 20 | J67..00 | Diseases of pancreas | 4 | 40 | 10.00 | 0.23 |

TABLE IV. THE POTENTIAL ADRS RELATED TO CANCER FOR ASPIRIN BASED ON PVALUE OF STUDENT'S T-TEST.

| Rank | Readcodes | Medical events | NB | NA | R1 | R2 |
|---|---|---|---|---|---|---|
| 1 | B33..00 | Other malignant neoplasm of skin | 85 | 334 | 3.93 | 1.95 |
| 2 | B76..00 | Benign neoplasm of skin | 76 | 224 | 2.95 | 1.31 |
| 3 | B34..00 | Malignant neoplasm of female breast | 19 | 99 | 5.21 | 0.58 |
| 4 | B22..00 | Malignant neoplasm of trachea, bronchus and lung | 14 | 80 | 5.71 | 0.47 |
| 5 | B46..00 | Malignant neoplasm of prostate | 57 | 163 | 2.86 | 0.95 |
| 6 | BB2..00 | [M]Papillary and squamous cell neoplasms | 15 | 86 | 5.73 | 0.50 |
| 7 | B8...00 | Carcinoma in situ | 16 | 58 | 3.63 | 0.34 |
| 8 | B13..00 | Malignant neoplasm of colon | 12 | 45 | 3.75 | 0.26 |
| 9 | B32..00 | Malignant melanoma of skin | 4 | 29 | 7.25 | 0.17 |
| 10 | 1J0..00 | Suspected malignancy | 11 | 51 | 4.64 | 0.30 |
| 11 | B14..00 | Malignant neoplasm of rectum, rectosigmoid junction and anus | 5 | 31 | 6.20 | 0.18 |
| 12 | BB3..00 | [M]Basal cell neoplasms | 7 | 37 | 5.29 | 0.22 |
| 13 | B49..00 | Malignant neoplasm of urinary bladder | 8 | 31 | 3.88 | 0.18 |
| 14 | B10..00 | Malignant neoplasm of oesophagus | 4 | 22 | 5.50 | 0.13 |
| 15 | B59..00 | Malignant neoplasm of unspecified site | 3 | 26 | 8.67 | 0.15 |
| 16 | BB5..00 | [M]Adenomas and adenocarcinomas | 19 | 54 | 2.84 | 0.32 |
| 17 | B17..00 | Malignant neoplasm of pancreas | 3 | 19 | 6.33 | 0.11 |
| 18 | B7F..00 | Benign neoplasm of brain and other parts of nervous system | 0 | 10 | 10.00 | 0.06 |
| 19 | B58..00 | Secondary malignant neoplasm of other specified sites | 3 | 18 | 6.00 | 0.11 |
| 20 | B43..00 | Malignant neoplasm of body of uterus | 2 | 15 | 7.50 | 0.09 |
| 21 | B57..00 | Secondary malig neop of respiratory and digestive systems | 3 | 17 | 5.67 | 0.10 |
| 22 | B81..00 | Carcinoma in situ of respiratory system | 1 | 10 | 10.00 | 0.06 |
| 23 | B82..00 | Carcinoma in situ of skin | 0 | 7 | 7.00 | 0.04 |
| 24 | BB4..00 | [M]Transitional cell papillomas and carcinomas | 6 | 19 | 3.17 | 0.11 |
| 25 | B44..00 | Malignant neoplasm of ovary and other uterine adnexa | 2 | 10 | 5.00 | 0.06 |
| 26 | B83..00 | Carcinoma in situ of breast and genitourinary system | 15 | 32 | 2.13 | 0.19 |
| 27 | B80..00 | Carcinoma in situ of digestive organs | 3 | 11 | 3.67 | 0.06 |
| 28 | B51..00 | Malignant neoplasm of brain | 1 | 7 | 7.00 | 0.04 |
| 29 | B11..00 | Malignant neoplasm of stomach | 5 | 14 | 2.80 | 0.08 |
| 30 | BBG..00 | [M]Fibromatous neoplasms | 0 | 4 | 4.00 | 0.02 |